# Transfer of spin to orbital angular momentum in the Bethe-Heitler process

Shaohu Lei,[1,2*] Shiyu Liu,[1,2*] Weiqing Wang,[1,2] Zhigang Bu,[1§] Baifei Shen[4] and Liangliang Ji,[1,3,†]

[1]*State Key Laboratory of High Field Laser Physics, Shanghai Institute of Optics and Fine Mechanics, Chinese Academy of Sciences, Shanghai 201800, China*
[2] *University of Chinese Academy of Sciences, Beijing 100049, China*
[3]*CAS Center for Excellence in Ultra-intense Laser Science, Shanghai 201800, China*
[4]*Shanghai Normal University, Shanghai 200234, China*

**Abstract:** According to the conservation of angular momentum, when a plane-wave polarized photon splits into a pair of electron-positron under the influence of the Coulomb field, the spin angular momentum (SAM) of the photon is converted into the angular momentum of the leptons. We investigate this process (the Bethe-Heitler process) by describing the final electron and positron with twisted states and find that the SAM of the incident photon is not only converted into SAM of the produced pair, but also into their orbital angular momentum (OAM), which has not been considered previously. The average OAM gained by the leptons surpasses the average SAM, while their orientations coincide. Both properties depend on the energy and open angle of the emitted leptons. The demonstrated spin-orbit transfer shown in the Bethe-Heitler process may exist in a large group of QED scattering processes.

*These authors contributed equally to this work.
§ zhigang.bu@siom.ac.cn
† jill@siom.ac.cn

## INTRODUCTION

Positrons can be generated through three main processes: the Bethe-Heitler (BH) process [1–6], the trident process [7–9], and the Breit-Wheeler (BW) process [10]. The Trident process is the direct interaction of electrons with the Coulomb field or laser field, while the BW process describes the generation of electron-positron pairs through collisions between photons, which implies that pure light can be turned into matter. The approach to produce abundant positrons employed in experiments is the BH process, where high-energy gamma photons split into electron-positron pairs under the influence of the Coulomb field of heavy nuclei. These gamma photons are usually emitted via bremsstrahlung of relativistic electrons with the Coulomb field. Recently, it has been demonstrated that laser-driven electrons can create copious positrons based on the BH process [11], which may have important applications in materials science, particle physics, laboratory astrophysics, and medical therapy [12–16].

In the BH process, polarization of the generated positrons can be preserved when the incident photon is circularly-polarized [17]. Here the spin angular momentum (SAM) of the incident photon is transferred to the polarization of the final electron-positron pair. The average polarization could reach 82% based on polarized electrons from storage rings [17]. In this work, we find that other than spin polarization, the lepton pairs could also gain non-negligible orbital angular momentum (OAM) in the BH process. We identify this effect by describing the final particles in the quantum vortex state, which inherently contains OAM along their axes of propagation. Theoretical description of the relativistic vortex electrons is based on the spinor Bessel mode [18,19] as an exact solution of the free Dirac equation in cylindrical coordinate. Recent interests in such particles are caused mainly by the magnitude of their total angular momentum (TAM) projection $\hbar m$ and the twisted electrons with $m \sim 1000$ can be routinely produced [20]. The generic features of vortex scattering not involving a specific process are analyzed in [21–24]. The properties of vortex states in various interaction processes have been explored in the interaction with atoms [25–27], the radiation process with twisted particles [28–33], the decay of the vortex muon [34], the features in external laser fields [35,36] and the spin-orbital coupling [18,37]. More comprehensive discussions can be found in several review articles on high-energy vortex states collisions [38] and free-electron vortex states [39]. Vortex electrons can trigger nuclear transitions that are not accessible by ordinary particles [40], which not only enhances the interaction rate but also provides an approach to identify the OAM of energetic particles.

Recently, it has been noticed that a vortex incident photon can transfer its OAM to the generated pair in the BH process [41]. Based on this study, we further find that even when the initial particles do not carry any vortex feature, the SAM of the incident photon converts into not only SAM but also OAM of the electron-positron pair. In other words, the electron-positron pair carry non-negligible OAM in polarized BH scattering. Such conversion is governed by the conservation of TAM [42]. Similar spin-to-orbital conversion has been revealed in the classical regime, where a circularly polarized plane wave laser pulse of relativistic intensity is focused and scattered by a plasma mirror [43–45]. The multi-photon absorption mechanism leads to a vortex light after reflection. Here in the QED regime, we obtain the distribution of the average OAM of the generated twisted positron in the BH process, the scattering probability with respect to the open angle, and further discuss the effects of incident photon energy and lateral collision distance.

**THEORY OF VORTEX POSITRON/ELECTRON**

We first introduce the description of the vortex (or twisted) electron/positron states. Here natural unit system $\hbar = c = 1$ and Gauss-Heaviside unit system where the fine-structure constant is $\alpha = e^2/4\pi \approx 1/137$ are applied in all calculations. Our convention for metric tensor takes the form $g^{\mu\nu} = diag(1,-1,-1,-1)$. We start from the negative-frequency solutions of Dirac equation

$$\psi(x) = v^s(p) e^{-i\mathbf{p}\cdot\mathbf{x} + iEt}, \tag{1}$$

where $v^s(p)$ is the Dirac bispinor of positron:

$$v^s(p) = \frac{1}{\sqrt{2(2\pi)^3}} \begin{pmatrix} \sqrt{1 - \frac{m_e}{E}} (\boldsymbol{\sigma} \cdot \mathbf{n}) \eta^s \\ \sqrt{1 + \frac{m_e}{E}} \eta^s \end{pmatrix}. \tag{2}$$

In Eqs. (1) and (2), $m_e$ is the mass of electron/positron, $E = \sqrt{m_e^2 + p^2}$ and $\mathbf{n} = \mathbf{p}/p$ are the total energy and the propagation direction of a positron, $\boldsymbol{\sigma}$ are the Pauli matrices and $\eta^s$ is the spin state in the rest frame of the particle, respectively. We consider the positron propagating along the quantization z-axis of the overall system. Using cylindrical coordinates $(r, \varphi, z)$ in coordinate space and $(p_\perp, \phi, p_z) = (p\sin\theta, \phi, p\cos\theta)$ in momentum space, the vortex positron state, which has well-defined values of the longitudinal linear momentum $p_z$, the modulus of the transverse momentum $|\mathbf{p}_\perp| = \varkappa$, and helicity $s$, is described by the wave function:

$$\psi_{\varkappa m p_z s}(\mathbf{x}) = \int d^2\mathbf{p}_\perp a_{\varkappa m}(\mathbf{p}_\perp) v^s(p) e^{-i\mathbf{p}\cdot\mathbf{x} + iEt}. \tag{3}$$

The Fourier spectrum of the negative-frequency Bessel beam [18] $a_{\varkappa m}(\mathbf{p}_\perp)$ in Eq(3) is:

$$a_{\varkappa m}(\mathbf{p}_\perp) = \frac{1}{\sqrt{2\pi}(-i)^m \varkappa} \delta(\varkappa - |\mathbf{p}_\perp|) e^{im\phi}. \tag{4}$$

By using the well-known integral relation $\frac{1}{2\pi} \int_0^{2\pi} e^{in\varphi + iz\cos\varphi} d\varphi = i^n J_n(z)$ ( $J_n(z)$ is the Bessel function of the first kind), we obtain the vortex wave function for positron:

$$\psi^s_{m,p_\perp,p_z}(x) = \frac{e^{-ip_z z + iEt}}{\sqrt{2}(2\pi)} \left[ \begin{pmatrix} \sqrt{1-\frac{m_e}{E}}\frac{p_z}{|p|}\sigma^3 \eta^s \\ \sqrt{1+\frac{m_e}{E}}\eta^s \end{pmatrix} J_m(\varkappa r) e^{im\varphi} + \frac{i\varkappa}{|p|}\sqrt{1-\frac{m_e}{E}} \begin{pmatrix} \begin{pmatrix} 0 & J_{m-1}(\varkappa r)e^{i(m-1)\varphi} \\ -J_{m+1}(\varkappa r)e^{i(m+1)\varphi} & 0 \end{pmatrix} \eta^s \\ 0 \end{pmatrix} \right]. \quad (5)$$

Such a Bessel state is the eigenmode of the TAM $J_z \psi^s_{m,p_\perp,p_z} = (m+s)\psi^s_{m,p_\perp,p_z}$, but not of the OAM and SAM separately [18]. Accordingly, the wave function of the twisted electron is obtained by substituting the positive frequency solution of the Dirac equation for negative frequency solution in Eq. (3).

## SCATTERING PROBABILITY OF THE VORTEX BH PROCESS

For comparison, we calculate the BH process in both the conventional plane-wave and the vortex scenarios, as shown in Fig. 1(a) and (b). Here the initial photon is circularly polarized and represented by a plane-wave packet in both cases. and the final electron-positron pair are represented by plane-wave state in the former and vortex states in the latter.

It should be noticed that the twisted states form the surface of a cone with the opening angle defined by $\theta = \arctan(\varkappa/p_z)$. The scattering matrix of Fig. 1(b) is then

$$\left(S_{fi}\right)_{wp} = \int \frac{d^3\mathbf{k}}{(2\pi)^3} \frac{1}{\sqrt{2\omega}} \rho(\mathbf{k}) 2\pi \delta(E_1 + E_2 - \omega) \mathcal{M}(k \to p_1 + \bar{p}_2), \quad (6)$$

where $\omega, k$ are the energy and momentum of the photon, and $E_1, p_1, E_2, p_2$ are the energy and momentum of the electron and positron, respectively. We assume that the photon wave packet has a Gaussian distribution with a central momentum $(0,0,\tilde{k}_z)(\tilde{k}_z > 0)$:

$$\rho(\mathbf{k}) = N \exp\left[\frac{-k_\perp^2 - (k_z - \tilde{k}_z)^2}{\tau^2}\right], \quad (7)$$

where $N = \left((2\pi/\tau)\sqrt{2/\pi}\right)^{3/2}$ is the normalization factor. Consider the narrow bandwidth approximation $\tilde{k}_z \gg \tau$:

$$|S_{fi}|^2 \approx C(\tau)(2\pi)\delta(E_1 + E_2 - \tilde{\omega})|\mathcal{M}(\tilde{k}_z \to p_1 + \bar{p}_2)|^2 \tag{8}$$

where

$$C(\tau) = \int \frac{d^3\mathbf{k}}{(2\pi)^3} \frac{d^3\mathbf{k}'}{(2\pi)^3} \left(\frac{2\pi}{\tau}\sqrt{\frac{2}{\pi}}\right)^3 \exp\left[\frac{-k_\perp^2 - (k_z - \tilde{k}_z)^2}{\tau^2}\right] \exp\left[\frac{-k_\perp'^2 - (k_z' - \tilde{k}_z)^2}{\tau^2}\right]^2 (2\pi)\delta(\omega - \omega')$$

$$= \frac{4\tilde{k}_z^2 \tau^2 + \tau^4 \left[1 - \exp\left(\frac{-2\tilde{k}_z^2}{\tau^2}\right)\right]}{8\pi \tilde{k}_z^2} \tag{9}$$

is the coefficient related to the wave packet and $\tilde{\omega}$ is the central value of energy. The scattering amplitude is $\mathcal{M}_{fi} = \mathcal{M}_1 + \mathcal{M}_2$ with

$$\mathcal{M}_1 = -ie^2 \int d^4x d^4x' \frac{d^4q}{(2\pi)^4} \bar{\psi}_{p_{1\perp},p_{1z}}^{+,m_1,s_1}(x) \gamma^\mu A_{k,\mu}^\lambda(x) \frac{(\slashed{q}+m_e)}{(q^2 - m_e^2)} e^{-iq\cdot(x-x')} \gamma^0 \psi_{p_{2\perp},p_{2z}}^{-,m_2,s_2}(x') A_0^{Coul}(x').$$

(10)

The Coulomb potential takes the form $A_0^{Coul}(\vec{x}) = -Ze/(4\pi|\vec{x}|) = -Ze\int 1/(2\pi)^3 \, e^{-i\vec{q}\cdot\vec{x}}/|\vec{q}|^2 d^3q$.

The wave function of twisted electron/positron can be deduced from (5). After integration we can get:

$$\mathcal{M}_{fi} = \mathcal{M}_{1,fi} + \mathcal{M}_{2,fi} = \frac{iZe^3}{4\sqrt{2\pi}} \sqrt{\frac{(E_1 - m_e)(E_2 - m_e)}{E_1 E_2}} \frac{1}{|p_1||p_2|} \delta(E_1 + E_2 - \omega) \xi^{s_1\dagger} \left( \Xi\Big|_{\substack{E_q = -E_2 = \omega - E_1 \\ q_z = p_{1z} - k_z = -p_{2z} - q_z'}} + \tilde{\Xi}\Big|_{\substack{E_q = E_1 \\ q_z = k_z - p_{2z}}} \right) \eta^{s_2} \tag{11}$$

with

$$\Xi^{\lambda}_{k_\perp,k_z} = \begin{pmatrix} \varsigma_{11}\delta_{\lambda,m_1-m_2} & \varsigma_{12}\delta_{\lambda,m_1-m_2+1} \\ \varsigma_{21}\delta_{\lambda,m_1-m_2-1} & \varsigma_{22}\delta_{\lambda,m_1-m_2} \end{pmatrix}. \tag{12}$$

The integer $\lambda$ represents the photon polarization, $\lambda=1$ represents right-handed circular polarization and $\lambda=-1$ represents left-handed. In addition, $m_1$ and $m_2$ are the OAM numbers of electrons and positrons, respectively. The matrix elements $\varsigma_{11}$, $\varsigma_{12}$, $\varsigma_{21}$ and $\varsigma_{22}$ are the sum of 9 terms respectively, similar to Ref. [41]. Each matrix element in $\Xi$ and $\tilde{\Xi}$ determines the creation probability of pairs with different spin polarizations, and the angular momentum (AM)-dependent Kronecker delta function gives the corresponding selection rule for the twisted BH process:

$$l = m_1 - m_2 + \Delta \tag{13}$$

with $\Delta=0,\pm1$. The minus sign before $m_2$ in Eq. (13) is consistent with the definition of positron AM. Finally, the scattering probability is:

$$\begin{aligned}
dp &= p_{1\perp}p_{2\perp}|S_{fi}|^2 dp_{1\perp}dp_{1z}dp_{2\perp}dp_{2z} \\
&= C(\tau)\frac{Z^2\alpha^3}{2\tilde{\omega}}\delta(\tilde{\omega}-E_1-E_2)\frac{p_{1\perp}p_{2\perp}(E_1-M)(E_2-M)}{E_1 E_2} \\
&\quad \times \frac{1}{|\mathbf{p}_1|^2|\mathbf{p}_2|^2}\text{Tr}\left[\xi^{s_1}\xi^{s_1\dagger}(\Xi+\tilde{\Xi})\eta^{s_2}\eta^{s_2\dagger}(\Xi+\tilde{\Xi})^\dagger\right]dp_{1\perp}dp_{1z}dp_{2\perp}dp_{2z}
\end{aligned} \tag{14}$$

**RESULTS**

We set the central energy and momentum of the photon wave packet as $\tilde{\omega}=5\text{MeV}$, $\tilde{k}_z=5\text{MeV}$, $\tilde{k}_\perp=0$, $\tau=0.004\text{MeV}$ and the photon polarization is $\lambda=1$. In Fig.1(c), we show the distribution of the scattering probability with respect to the polar angle. One notices that the channel with $s_2=1/2$ is significantly higher than that of $s_2=-1/2$ in both cases, suggesting that the positron is highly polarized. In other words, the polarization of the incident photon is efficiently transferred to the electorn-positron pair. The total polarization reaches 41.82%, which is less than the photon SAM h-bar. As shown later, this is due to the spin-orbit transfer in the BH process.

The scattering probability exhibits similar distribution pattern in both scnearios, where the maximum value appears along the direction close to the z-axis $\theta \sim 11°$. The probability peaks even closer to the propagation direction when the incident photon is more energetic. On the other hand, the peak and overall probabilities of vortex scenario is slightly smaller than that of the pure plane-wave case, as shown in Fig. 1(c). The lower dashed lines refers to a sum of $m_2$ from -1 to 1 and the other from -4 to 4. It is due to the fact that a plane wave can unfold into a superposition of infinite vortex states. As the range for summing the OAM number $m_2$ increases, the overall probability of the vortex scattering gradually converges to the plane-wave case. Fig.1(d) shows the scattering probability as a function of the positron energy, where consistency is observed between both scenarios. We see that in the case of $s_2 = 1/2$, the probability peaks in the high-energy section, around 3.34 MeV. This is also the dominating channel. On the contrary, the lesser channel reaches maximum in the low-energy part 1.41 MeV. The situation reverses when switching the photon polarization. This suggests that the spin and energy of the generated particles are strongly correlated. The peak of the total probability is around 2.5 MeV, which is exactly half the energy of the incident photon. Fig.1(d) also reveals the spin polarization property for positron. In low-energy region, the polarization is small and becomes larger as the energy increase. Similar trend can be found in [46] where the transfer of longitudinal spin in multiphoton BH pair production has been analyzed, and in [47] which analyzes the spin effect in the BW process.

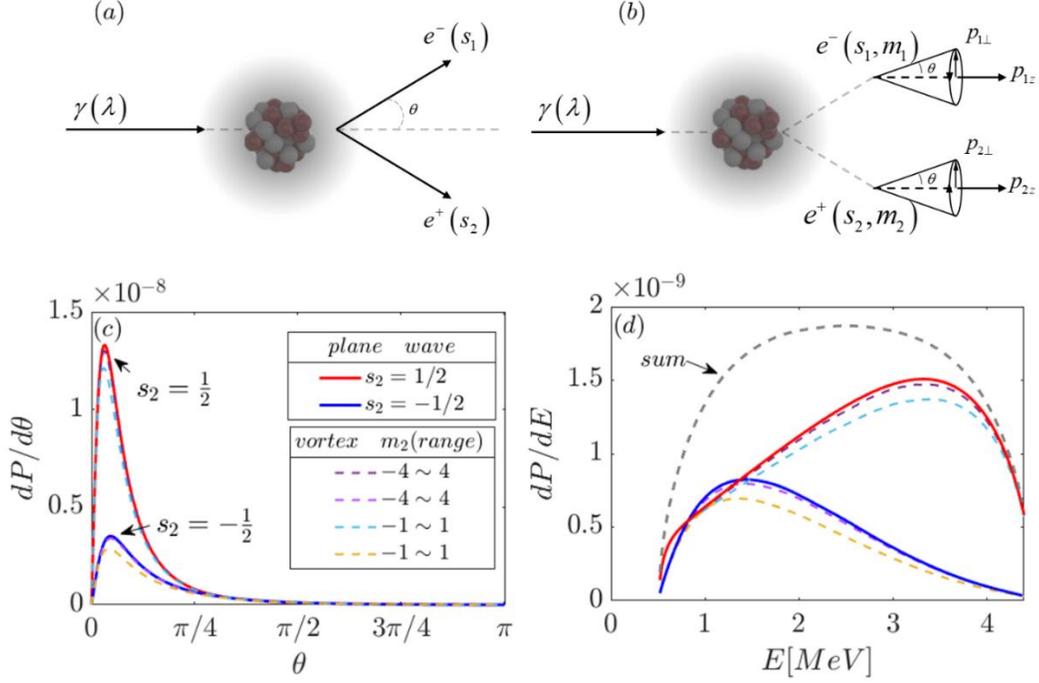

**Figure 1.** Schematic diagram of the Bethe-Heitler process with electron-positron pair described in the plane-wave state (a) and vortex state (b), from a circularly polarized plane-wave photon bombarding a Cu nucleus. (c) Distribution of the scattering probability with respect to the polar angle (c) and energy of the positron (d) in both scenarios, in the case of $s_2=1/2$ (red solid, dashed) and $s_2=-1/2$ (blue solid, dashed). The dashed lines represent the range of sum over the $m_2$ value.

In the following, we present the scattering probability as a function of the OAM number of the electron-positron pair. It can be seen from Fig. 2 that the maximum value of the spectrum is at $m=0$, accounting for about 58% of the total probability. When $\lambda=1$, the sub-maximum value of the electron is at $m=1$, while the positron at $m=-1$. It indicates that the electron tends to gain net OAM of $m>0$ and the positron gains $m<0$.

We average the spectra in the range of $m_2 \sim [-5,5]$. The average values of OAM and SAM are $\bar{m}_1 = 0.2898$ and $\bar{s}_1 = 0.2091$ for the electron and $\bar{m}_2 = -0.2899$, $\bar{s}_2 = -0.2091$ for the positron. It is quite surprising that the absolute value of OAM is notably higher than the SAM for both particles, which cannot be seen using the conventional plane-wave calculation. These values satisfy the conservation of TAM $\bar{m}_1 - \bar{m}_2 + \bar{s}_1 - \bar{s}_2 = \lambda$. When $\lambda$ is switched from 1 to -1, the OAM spectra of positron and electron are also interchanged, as seen in Fig. 2(b).

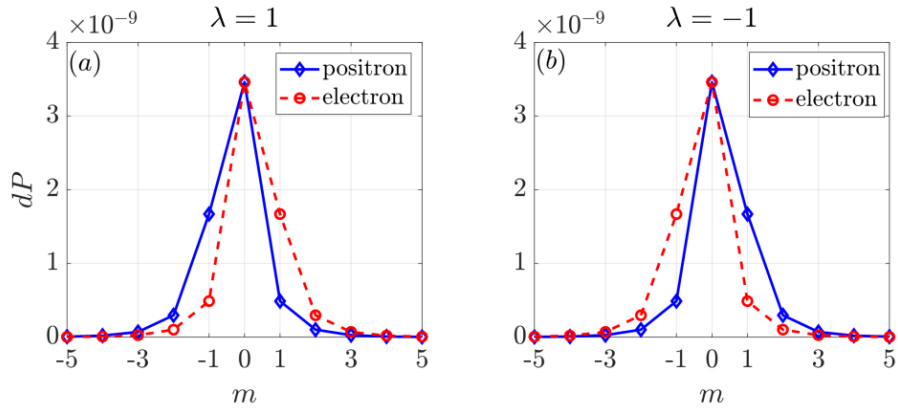

**Figure 2**. The OAM spectra of generated positron (blue) and electron(red) corresponding to the photon polarization of (a) $\lambda=1$ and (b) $\lambda=-1$.

The averaged OAM and SAM distributions are further elaborated at different energies and opening angles in Fig.3. It can be seen in Fig. 3(c) that higher average spin appears in the high energy region, albeit the spin orientation with respect to the momentum flips in forward scattering ($s < 0$ for $\theta < \pi/2$) and backward scattering ($s > 0$ for $\theta > \pi/2$). This trend is almost reversed towards the low energy end, where the sign is positive/negative in the forward/backward direction. On the other hand, the OAM stays $m < 0$ overall. The average OAM value peaks in the high energy region for back-scattered positron. Another narrow peak appears at an opening angle where the probability maximizes, denoted by the angle obtained in Fig. 1(c). Again, when the photon spin flips the value of both average angular momentum is reversed, as seen in Fig. 3(b) and (d). In most parameter regions in Fig. 3, we have $\bar{m}+\bar{s}=-0.5$ which is the half of the polarization of photon $\lambda$, and the extra minus sign is consistent with the definition of positron AM.

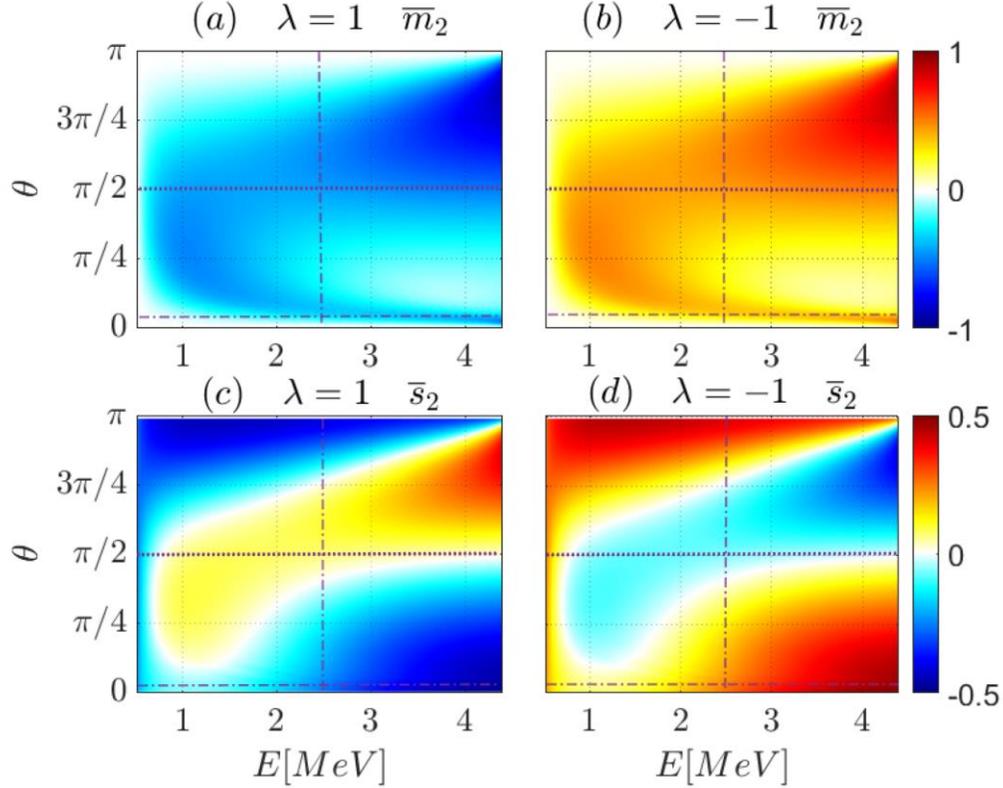

**Figure 3.** Distribution of the averaged OAM (a) & (b) and averaged SAM (c) & (d) of the produced positrons as a function of energy and angle. The incident photon energy is 5MeV. The red dotted line in the figure represents where the probability maximum occurs. The vertical line at $\pi/2$ is to distinguish between forward and backward scattering.

## DISCUSSION

We then considered the effect of different incident photon energies and lateral impact parameter **b** on the process. In Fig. 4(a) we can see that the OAM spectra of positron are similar. When the polarization of photon is $\lambda=1$, the probability at $m_2=-1$ is notably higher than that at $m_2 = 1$, indicating net avarage OAM deviating from $m_2 = 0$. The probability of other vortex states is greatly reduced. For higher photon energy, the total scattering probability becomes larger, and the spectral width is broadened. This suggests that vortex states of higher orders can be generated more efficiently by increasing the photon energy. In Fig. 4(b), as the photon energy increases, the average OAM $\bar{m}_2$ shows insignificant variation, which is around -0.25. This shows that the OAM obtained by the positron is comparable or slightly higher than that of the SAM.

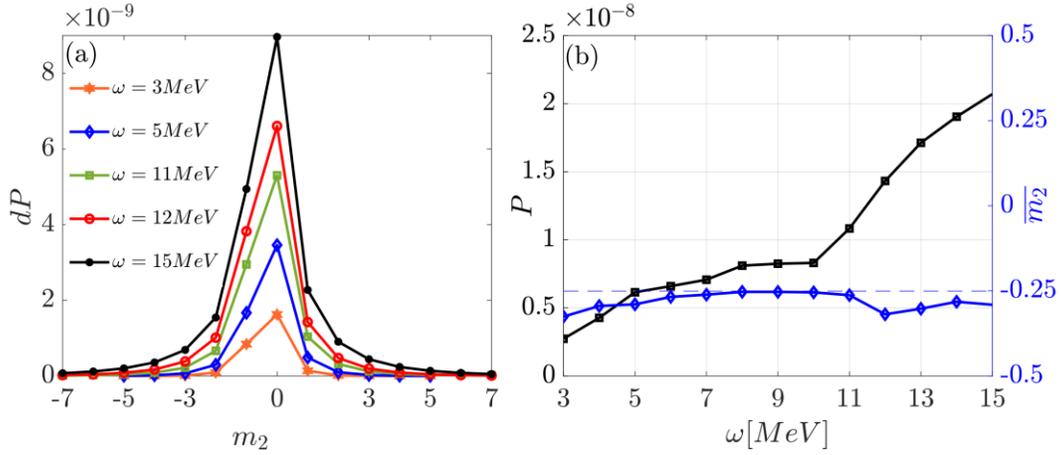

**Figure 4.** (a) OAM spectra of the positron corresponding to different incident photon energies. (b) Distribution of the total scattering probability and average OAM versus incident photon energy.

The above derivation assumes that the incident photon strikes perfectly at the center of the nucleus along the z-axis (impact parameter $b = 0$). In the following, we consider when there is a certain deviation from the z-axis during the collision. The corresponding wave function of the incident photon in the momentum representation can be expressed in which the axis is shifted in the transverse plane by a distance b from the potential center, [26]:

$$\rho(\mathbf{k},\ \mathbf{b}) = N\exp\left[\frac{-k_\perp^2 - \left(k_z - \tilde{k}_z\right)^2}{\tau^2}\right]e^{-i\mathbf{k}_\perp \mathbf{b}}. \tag{15}$$

In the numerical simulation, we set $\tau = 0.004 MeV, \tilde{k}_z = 5 MeV$, fulfilling the narrow bandwidth approximation. Then we get

$$\begin{aligned}
\left|S_{fi}\right|^2 &= \int \frac{d^3\mathbf{k}}{(2\pi)^3}\frac{d^3\mathbf{k}'}{(2\pi)^3}\rho(\mathbf{k},\ \mathbf{b})\rho(\mathbf{k}',\ \mathbf{b}) \\
&\quad \times (2\pi)\delta(\omega - \omega')(2\pi)\delta(E_1 + E_2 - \tilde{\omega})\left|\mathcal{M}(k_z, k_\perp \to p_1 + \bar{p}_2)\right|^2 \\
&\approx \int \frac{d^3\mathbf{k}}{(2\pi)^3}\frac{d^3\mathbf{k}'}{(2\pi)^3}\rho(\mathbf{k},\ \mathbf{b})\rho(\mathbf{k}',\ \mathbf{b}) \\
&\quad \times (2\pi)\delta(\omega - \omega')(2\pi)\delta(E_1 + E_2 - \tilde{\omega})\left|\mathcal{M}(\tilde{k}_z, \tilde{k}_\perp \to p_1 + \bar{p}_2)\right|^2 \\
&= C(\mathbf{b})(2\pi)\delta(E_1 + E_2 - \tilde{\omega})\left|\mathcal{M}(\tilde{k}_z, \tilde{k}_\perp \to p_1 + \bar{p}_2)\right|^2
\end{aligned} \tag{16}$$

Because $\tilde{k}_\perp$ equals zero, $\left|\mathcal{M}(\tilde{k}_z, \tilde{k}_\perp \to p_1 + \bar{p}_2)\right|^2$ is independent to the rest of the integral. Therefore, the impact parameter $\mathbf{b}$ only affects the magnitude of the scattering probability, but not the OAM and SAM distribution patterns. With the formula: $e^{-i\mathbf{k}_\perp \mathbf{b}} = \sum_n i^{-n} J_n(k_\perp b) e^{-in\phi_b + in\phi_k}$, the term taking into account of the impact parameter is

$$\begin{aligned}
C(\mathbf{b}) &= \int \frac{d^3\mathbf{k}}{(2\pi)^3}\frac{d^3\mathbf{k}'}{(2\pi)^3}\left(\frac{2\pi}{\tau}\sqrt{\frac{2}{\pi}}\right)^3 \exp\left[\frac{-k_\perp^2 - \left(k_z - \tilde{k}_z\right)^2}{\tau^2}\right]e^{-i\mathbf{k}_\perp \mathbf{b}}\exp\left[\frac{-k_\perp'^2 - \left(k_z' - \tilde{k}_z\right)^2}{\tau^2}\right]e^{i\mathbf{k}_\perp' \mathbf{b}} \\
&\quad \times (2\pi)\delta(\omega - \omega') \\
&= \exp\left(-\frac{2\tilde{k}_z^2}{\tau^2}\right)\left(\frac{1}{\tau}\sqrt{\frac{2}{\pi}}\right)^3 \int_0^{+\infty} d\omega\omega^4 \exp\left(-\frac{2\omega^2}{\tau^2}\right)\left[\int_{-1}^{1} d\cos\theta \exp\left(\frac{2\tilde{k}_z\omega\cos\theta}{\tau^2}\right)J_0(\omega b \sin\theta)\right]^2
\end{aligned} \tag{17}$$

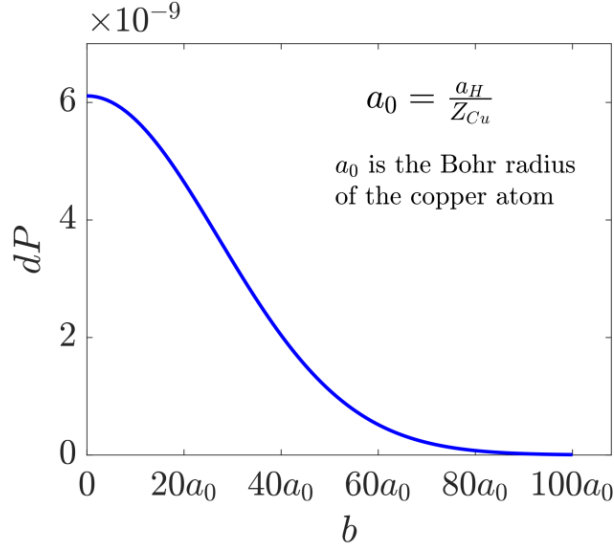

**Figure 5.** Scattering probability versus the collision distance b. Here $a_0$ is the Bohr radius of the copper atom.

In Fig. 5, we see that as the lateral distance increases, the scattering probability becomes smaller. At around $|b|=30a_0$, the scattering probability is reduced to half that of a concentric collision where $a_0$ is the Bohr radius of Cu. When the lateral distance is larger than $100a_0$, the probability of the BH process occurring can be ignored. We emphasize that this is only the case when the narrow bandwidth approximation is applied. When this requirement is not valid, the effect of the parameter b cannot be extracted from the matrix M. The results will be much more complicated.

## CONCLUSION

We calculated the BH process using the initial state of the plane wave packet and the final state of the vortex state. The OAM distribution of produced positrons and the distribution of scattering probability with angle are numerically simulated. It is theoretically proved that the spin angular momentum of the plane wave photons can be converted not only to the SAM of the final positron-electron pair, but also to their OAM.

Finally, we discussed the cases of incident photons with different energies and non-concentric collisions. With the increase of incident photon energy, the scattering probability increases, and with the increase of the lateral collision distance, the scattering probability decreases. This work deals with the angular momentum of BH process along the photon propagation direction. It is helpful for understanding angular momentum conversion and spin-orbit coupling effects. Since the initial photon is in the plane-wave state rather than vortex state, the vortex effect revealed here may exist in many other scattering processes.

**ACKNOWLEDGMENTS**

This work is supported by the National Science Foundation of China (Nos. 11875307 and 11935008), the Strategic Priority Research Program of Chinese Academy of Sciences (Grant No. XDB16010000) and the Ministry of Science and Technology of the People's Republic of China (Grant Nos. 2018YFA0404803).